# Performance Analysis of Quantitative Phase Retrieval Method in Zernike Phase Contrast X-ray Microscopy

Heng Chen (陈恒)[1,2], Kun Gao (高昆)[2], Dajiang Wang(王大江)[2], Li Song (宋礼)[2], and Zhili Wang (王志立)[2; 1)]

[1] Department of Physics, University of Science and Technology of China, Hefei, Anhui, 230026, China

[2] National Synchrotron Radiation Laboratory, University of Science and Technology of China, Hefei, Anhui, 230029, China

[1)] Correspondence email: wangnsrl@ustc.edu.cn

**Abstract:** Since the invention of Zernike phase contrast method in 1930, it has been widely used in optical microscopy and more recently in X-ray microscopy. Considering the image contrast is a mixture of absorption and phase information, we recently have proposed and demonstrated a method for quantitative phase retrieval in Zernike phase contrast X-ray microscopy. In this contribution, we analyzed the performance of this method at different photon energies. Intensity images of PMMA samples are simulated at 2.5 keV and 6.2 keV, respectively, and phase retrieval is performed using the proposed method. The results demonstrated that the proposed phase retrieval method is applicable over a wide energy range. For weakly absorbing features, the optimal photon energy is 2.5 keV, from the point of view of image contrast and accuracy of phase retrieval. On the other hand, in the case of strong absorption objects, a higher photon energy is preferred to reduce the error of phase retrieval. Those results can be used as guidelines to perform quantitative phase retrieval in Zernike phase contrast X-ray microscopy with the proposed method.

## 1 Introduction

In 1930, Zernike solved the problem of poor absorption contrast in optical microscopy by introducing a quarter-wavelength shift in the relative phase between specimens diffracted and undiffracted waves [1]. It is obvious that the absorption for X-rays is typically weaker than that for visible light. So X-ray microscopy can greatly profit from the Zernike phase contrast [2-5]. The first reported results were concerned for soft X-rays in 1995 [2]. Recently, the research activity is shifted to the intermediated and hard X-ray region [6-15]. A 20 nm spatial resolution has been achieved at a photon energy of 8 keV [16].

For an X-ray microscope working in the Zernike phase contrast mode, the measured image intensity contains a mixture of absorption and phase information. So, for the interpretation of the image, it is necessary to perform quantitative phase retrieval. Some approaches for phase retrieval have been reported by Yin et al. [17] and Liu et al. [9]. However, these methods require a series of sample's intensity images, and multiple exposures may result in excessive radiation dose to biological samples. Recently, we proposed a novel method for quantitative phase retrieval in X-ray Zernike phase contrast microscopy [18]. The novelty of this method is the use of two different intensity images, which were measured with a $\pi/2$ and a $3\pi/2$ phase ring, respectively. These two intensity images are then proceed by the phase retrieval method, to yield the phase shift of the feature of interest. Compared to previous approaches, this method only requires two intensity images, which will simplify the experimental procedure and increase the imaging speed. Furthermore, the numerical experiment results demonstrated that this method is effective for quantitative phase retrieval for both weakly and strongly absorbing objects, at a photon energy of 2.5 keV.

In this contribution, we analyzed the applicability of the developed phase retrieval method at different photon energies. We numerically simulate Zernike phase contrast X-ray microscopes at photon energies of 2.5 keV and 6.2 keV, respectively. The two intensity images required by the method are calculated, and phase retrieval is performed for different samples. The presented results demonstrate that the phase retrieval method is feasible over a wide energy range. Besides, it is shown that for those low-absorption-contrast features, intermediate X-ray microscopy around 2.5 keV is optimal from the point of view of image contrast and accuracy of phase retrieval. In the case of strongly absorbing features, a higher photon energy is preferred to reduce the error in phase retrieval. The obtained results can serve as guidelines in experimental researches for various samples.

**2 Method**

In a Zernike phase contrast X-ray microscope, a capillary reflective condenser is typically used to produce a hollow cone illumination; the light passing through the sample is separated into an undiffracted and a diffracted part, which carries the information of the sample. The undiffracted radiation goes through the phase ring, which sets at the back focal plane of the zone plate objective. In the detect plane, the undiffracted part interferes with the diffracted part. Eventually, the phase shifts of the sample are translated into intensity modulations detectable in the collected images.

The key point of the quantitative phase retrieval method is that two phase rings are successively inserted into the radiation path; hence, two measurements of the intensity images can be performed in sequence in the detector plane. Moreover, with the measured intensity images, quantitative phase retrieval of the sample can be achieved in X-ray Zernike phase contrast microscopy. The quantitative phase retrieval method is given by Eq. (1), and detailed mathematical derivations can be found in our previous work [18].

$$\varphi = \arcsin\left[2B/(B^2+1)\right]. \qquad (1)$$

with $B = \left(a_1^2 I_1' - a_2^2 I_2' - a_1^2 + a_2^2\right) / \left\{a_1 a_2 \left[a_1 I_1' + a_2 I_2' - (a_1 + a_2)\right]\right\}$, where $a_1 = \exp(-2\pi\beta_p t_{p1}/\lambda)$, $a_2 = \exp(-2\pi\beta_p t_{p2}/\lambda)$, $\beta_p$ is the imaginary part of the refractive index of the phase ring material. $t_{p1}$ and $t_{p2}$ are the thickness of the π/2 and 3π/2 phase ring, respectively, and λ is the X-ray wavelength. $I_1'$ represents the normalized intensity image corresponding to a π/2 phase ring and $I_2'$ represents the normalized intensity image when using a 3π/2 phase ring.

Eq. (1) summarizes the quantitative phase retrieval method reported in our previous work. Moreover, the method has been validated of both weak and strong absorption features by numerical experiments [18]. In the following sections, we will discuss the applicability of the quantitative retrieval method at various photon energies, and summarize the suitable conditions for future experimental research.

**3 Result and discussion**

Numerical experiments have been performed to discuss the applicability of the developed phase retrieval method at different photon energies. We numerically simulated X-ray Zernike phase contrast microscopes at photon energies of 2.5 keV and 6.2 keV, respectively. Fresnel zone plates of Au of 400 nm thick with 30 nm outmost zone width were used as the objective to reach a spatial resolution better than 30 nm. We used two phase rings made of nickel, where one is π/2-phase shifting and the other is a 3π/2-phase shifting. In the simulation, Poisson distributed shot noise has been added in the measured images.

The first sample for test is a representative of weakly absorbing objects. The sample is a zone plate made of polymethyl methacrylate (PMMA) on silicon nitride [17], and the thicknesses are 1000 nm, 800 nm, 600 nm, 400 nm, 200 nm, 200 nm, going from the inner to the outer rings, respectively. The refractive index of the test sample can be found in references [19]. In the following, we simulate X-ray Zernike phase contrast microscopes for the imaging of the weakly absorbing sample at photon energies of 2.5 keV and 6.2 keV, respectively.

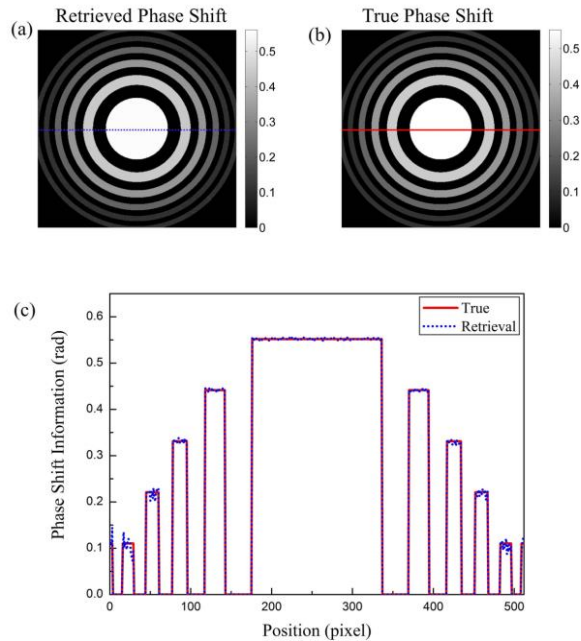

Fig. 1. (a) and (b) show the retrieved and the true phase shift of the test sample at a photon energy of 2.5 keV, respectively. (c) is the line profile comparison between the retrieved and the true phase shift.

For a photon energy of 2.5 keV, the results of the test sample are shown in Fig. 1. The retrieval result of the test sample is shown in Fig. 1(a). And Fig. 1(b) presents the true phase shift. As can be seen, Figs. 1(a) and 1(b) match qualitatively well. Additionally, a line profile comparison has been also performed, as shown in Fig. 1(c), and a quantitative agreement is achieved. Going from the inner to the outer rings, the true phase shifts of sample are 0.551785 rad, 0.441428 rad, 0.331071 rad, 0.220714 rad, 0.110357 rad, 0.110357 rad, respectively; and the corresponding retrieved phase shifts are 0.552127 rad, 0.442487 rad, 0.331279 rad, 0.218562 rad, 0.103971 rad, 0.103971 rad, respectively. Note that these values are quantitatively in good agreement. The small mismatch in the region of the outer ring is due to the added shot noise, and the average value of the retrieved phase shift matches well with the true phase shift. Results shown in Fig. 1 confirm the feasibility of our method for weakly absorbing objects at a photon energy of 2.5 keV.

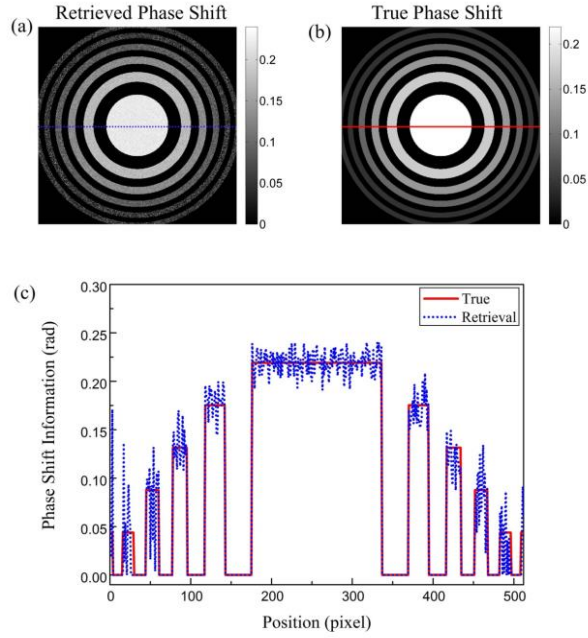

Fig. 2. (a) and (b) show the retrieved and the true phase shift of the test sample at a photon energy of 6.2 keV, respectively. (c) is the line profile comparison between the retrieved and the true phase shift.

With the photon energy increased to 6.2 keV, the results of the first test sample are shown in Fig. 2. The retrieved phase shift of the sample by use of Eq. (1) is shown in Fig. 2(a). For a comparison, Fig. 2(b) shows the true phase shift, which visually agrees with Fig. 2(a). For a quantitative comparison, Fig. 2(c) shows the line profile comparison has also been performed, and the results is shown in Fig. 2(c). Going from the inner to the outer rings, the true phase shifts of sample are 0.219052 rad, 0.175242 rad, 0.131431 rad, 0.0876208 rad, 0.0438104 rad, 0.0438104 rad, respectively; and the retrieved phase shifts are 0.21993 rad, 0.176702 rad, 0.12719 rad, 0.093412 rad, 0.050014 rad, 0.050014 rad, respectively. Note that these values are quantitatively in agreement. Owing to the added shot noise, the retrieved phase shift mismatches the true phase shift when the thickness of the sample is smaller than 400 nm. Anyway, the results shown in Fig. 2 confirm the feasibility of our method for weak absorption objects at a photon energy of 6.2 keV.

Secondly, we consider the case of a strongly absorbing object. Following our previous work [18], we again used a Siemens star like, which consisted of copper features surrounded by water as the second test sample. The sample mimics a high absorbing system in a cellular environment with a total thickness of 5 μm. As the feature of interest, the thicknesses of the copper range from 30 nm at the 3 o'clock position to 105 nm with 5 nm steps and structures are arranged anticlockwise with a fan angle of π/16.

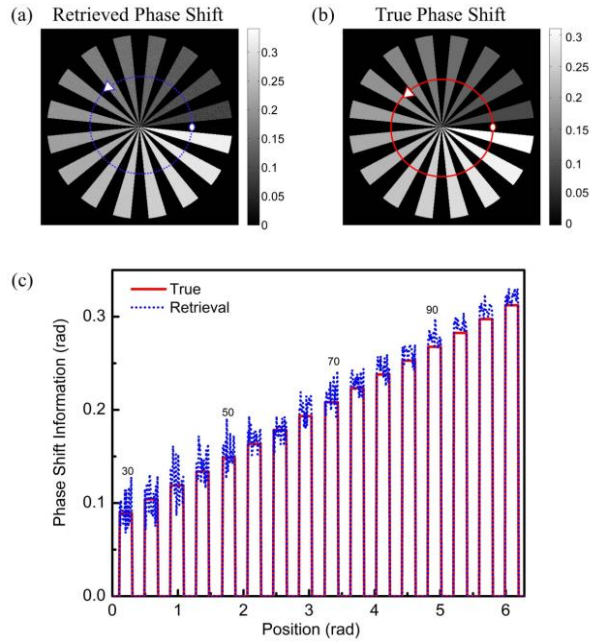

Fig. 3. (a) and (b) show the retrieved and the true phase shift of the second test sample at a photon energy of 2.5 keV, respectively. (c) is the profile comparison between the retrieved and the true phase shift. The numbers on figure (c) (30, 50, 70 and 90) indicate the thicknesses of the Copper features.

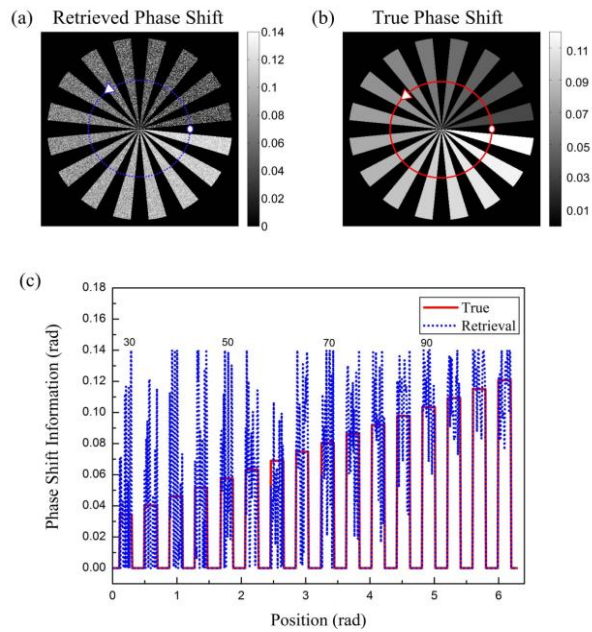

Fig. 4. (a) and (b) show the retrieved and the true phase shift of the second test sample at a photon energy of 6.2 keV, respectively. (c) is the profile comparison between the retrieved and the true phase shift. The numbers on figure (c) (30, 50, 70 and 90) indicate the thicknesses of the Copper features.

The results of the strongly absorbing sample are shown in Figs. 3 and 4 at different photon energies. The phase shift of the second test sample is retrieved by use of Eq. (1), and the results are shown in Figs. 3(a) and 4(a), corresponding to the photon energies of 2.5 keV and 6.2 keV, respectively. By contrast, Figs. 3(b) and 4(b) are the true phase shift of the sample at photon energies of 2.5 keV and 6.2 keV, respectively. Besides, we present quantitative comparisons of the line profile in Figs. 3(c) and 4(c), respectively. As can be seen in Fig. 3(c), when the thickness is less than 35 nm the retrieval result is affected by the added shot noise; while the thickness of Cu ranges from 40 to 80 nm, the retrieved phase shift well match the true value. But when the thickness is larger than 80 nm, the retrieved phase shift is substantially larger than the true value. However, with the increase of the photon energy, the linear approximation holds well and the retrieved phase shift matches well with the true phase shift. This conclusion can be proved by Fig. 4(c); it is obvious that it has a good agreement between the retrieved result and the true value, when the thickness of Cu is larger than 80 nm. Images and analysis confirm the reliability of the proposed method for strongly absorbing objects at a proper photon energy.

From Fig. 1 to Fig. 2, we know that the quantitative phase retrieval method is feasible in X-ray Zernike phase contrast microscope at different photon energies. By comparing Fig. 1(c) with Fig. 2(c), it is obvious that when the photon energy is 2.5keV, a Zernike phase contrast microscope may provide a good image contrast for the PMMA sample and the retrieved phase shift matches well with the true phase shift. Therefore, it is the optimal energy for the biological specimens, typically with low-absorption-contrast features, in X-ray Zernike phase contrast microscopy. In addition, for strong-absorption-contrast features, the liner approximation is poor at a low photon energy, which results in a large error in the retrieved phase shift. However, with increased photon energy, the linear approximation holds well and the retrieved phase shift fits the true values well. As shown in Fig. 4(c), the retrieved phase shift of the sample matches well the true values, when the thickness of Cu is 80 to 100 nm at a photon energy of 6.2 keV. Therefore, in a real experiment, one may estimate the attenuation of the investigated object before imaging and determine suitable incident X-ray energy. Also, the prior knowledge of the object absorption can be used to correct the results of the phase retrieval.

## 4 Conclusion

In conclusion, we have demonstrated the feasibility of the phase retrieval method in X-ray Zernike phase contrast microscopy at different photon energies. A good agreement between retrieved results and true values is obtained at intermediate and hard X-ray energies. These presented results show that for low-absorption-contrast features, the optimal photon energy may be 2.5 keV from point of view of image contrast and accuracy of phase retrieval. Furthermore, the quantitative method is also feasible for strong-absorption-contrast features by increasing the incident X-ray energies properly. With the development of the technology, the proposed method will match the instrument nowadays used at synchrotron radiation facilities. And thanks to the development of X-ray optics, we foresaw that this method could be successfully implemented in biomedical and materials science research.

**Acknowledgements**

This work was supported by the States Key Project for Fundamental Research (2012CB825801), the Science Fund for Creative Research Groups (11321503), the National Natural Science Foundation of China (11475170, 11205157 and 11179004) and the Anhui Provincial Natural Science Foundation (1508085MA20).